\documentclass[twocolumn,showpacs,preprintnumbers,amsmath,amssymb]{revtex4}
\usepackage{graphicx}% Include figure files
\usepackage{dcolumn}% Align table columns on decimal point
\usepackage{bm}% bold math
\input{epsf.tex}

\begin{document}

\title{Evidence for  narrow resonant structures at $W \approx 1.68$ and $W \approx 1.72$ GeV  
in real Compton scattering off the proton}

\author{V.~Kuznetsov$^{1,*}$}
\author{F.Mammoliti$^{2,3}$}
\author{V.Bellini$^{2,3}$}
%\author{M. V.~Polyakov$^{1,4}$}

%\author{H.-S.Dho$^{1}$}
\author{G.Gervino$^{4,5}$}
\author{F.Ghio$^{6,7}$}
%\author{A.Giusa$^{2,3}$}
\author{G.Giardina$^{8}$}
%\author{A.~Kim$^{1,12}$}
\author{W.~Kim$^{9}$}
\author{G.~Mandaglio$^{2,8}$}
\author{M.L.~Sperduto$^{2,3}$}
\author{C.M.~Sutera$^{2,3}$}
\vspace{0.5 cm}

\affiliation{$^1$Petersburg Nuclear Physics Institute, 188300 Gatchina, Russia}
\affiliation{$^{2}$INFN - Sezione di Catania, via Santa Sofia 64, I-95123 Catania, Italy}
\affiliation{$^{3}$Dipartimento di Fisica ed Astronomia, Universit\'a di Catania, I-95123 Catania, Italy}
\affiliation{$^{4}$Dipartamento di Fisica Sperimentale, Universit\'a di Torino, 
via P.Giuria, I-00125 Torino, Italy}
\affiliation{$^{5}$ INFN - Sezione di Torino, I-10125 Torino, Italy}
\affiliation{$^{6}$INFN - Sezione di Roma, piazzale Aldo Moro 2, I-00185 Roma, Italy}
\affiliation{$^{7}$Instituto Superiore di Sanit\'a, viale Regina Elena 299, I-00161 Roma, Italy}
\affiliation{$^{8}$Dipartimento di Fisica e di Scienze della Terra - Universit\`a di Messina,
salita Sperone 31, 98166 Messina, Italy}
\affiliation{$^{9}$Kyungpook National University, 702-701 Daegu, Republic of Korea}
\date{\today}

\begin{abstract}

First measurement of the beam asymmetry $\Sigma$ for Compton scattering off the proton
in the energy range $E_{\gamma}=0.85 - 1.25$ GeV is presented.
The data reveals two narrow structures at $E_{\gamma}= 1.036$ and $E_{\gamma}=1.119$ GeV. 
They may signal narrow resonances with masses near $1.68$ and $1.72$ GeV,
or they may be generated by the sub-threshold $K\Lambda$ and $\omega p$ production.  
Their decisive identification requires additional theoretical and experimental efforts.

\pacs{14.20.Gk,13.60.Rj,13.60.Le}
\end{abstract}
\thanks{Electronic address: Slava@pnpi.spb.ru}
\maketitle

The observation of a narrow enhancement at $W \sim 1.68$ GeV
in $\eta$ photoproduction~\cite{gra,kru,kru2,kas,mainz1} and
Compton scattering off the neutron~\cite{comp} (the so-called ``neutron anomaly") is of
particular interest because it may signal a nucleon resonance with unusual properties:
a mass near $M\sim 1.68$~GeV, a
narrow ($\Gamma \leq 25$  MeV) width, a strong photoexcitation on the neutron,
and a suppressed decay to $\pi N$ final state~\cite{az,tia,kim,arndt,mart}.
Such resonance was never predicted by the traditional Constituent Quark Model~\cite{cqm}. 
On the contrary, its properties coincide
surprisingly well with those expected for an exotic state predicted in the framework of the
chiral soliton model~\cite{dia,max,dia1,michal,mfa}.

On the other hand, several groups~\cite{int}
explained the bump in the $\gamma n \to \eta n$ cross section in terms of
the interference of well-known wide resonances.
Although this assumption was challenged by the results on Compton scattering off the neutron~\cite{comp} 
(this reaction is governed by different resonances),  it is widely discussed in literature.  
Another explanation was proposed by M.~Doring and K.~Nakayama~\cite{dor}. They explained the neutron anomaly
as virtual sub-threshold $K \Lambda$ and $K \Sigma$ photoproduction (``cusp effect").
At present, the decisive identification of the narrow peculiarity at $W\sim 1.68$ GeV is a challenge for both
theory an experiment.  

One benchmark signature of the $N^*(1685)$ resonance (if it does exist) is strong photoexcitation on the 
neutron and weak (but not zero) photoexcitation on the proton. Such resonance
would appear in cross section on the proton as a minor peak(dip) structure which might 
be not (or poorly) seen in experiment.  However its signal may be amplified in polarization observables
due to the interference with other resonances.
 
The recent high-precision and high-resolution measurement of the $\gamma p \to \eta p$ cross section 
by the A2@MaMiC Collaboration~\cite{mainz2} made it possible to retrieve a small dip at $W\approx 1.69$ GeV
which was not resolved in previous experiments.
At the same time the revision of the GRAAL beam asymmetry $\Sigma$ for
$\gamma p \to \eta p$ revealed a resonant structure
at $W=1.685$ GeV~\cite{acta,jetp} (see also~\cite{an1}).
The bump in the Compton scattering off the neutron at $W=1.685$ GeV was observed
at GRAAL~\cite{comp}. The motivation for this work was to search, in analogy with $\eta$ photoproduction,
a resonant structure in polarization observables for Compton scattering on the proton.

In this Rapid Communication, we report on the first measurement of the beam asymmetry $\Sigma$ for
Compton scattering off the proton in the range of incident-photon energies
$E_{\gamma}=0.85 - 1.25$~GeV.
The data were collected at the GRAAL facility~\cite{pi0} from 1998 to 2003 in a number of 
data-taking periods. The main difference between the periods was the usage of either UV or
green laser light. A highly-polarized and tagged photon beam 
was produced by means of backscattering of this light on 6.04~GeV electrons
circulating in the storage ring of the European Synchrotron Radiation Facility (ESRF,Grenoble, France).
The tagged photon-energy range was $\sim 0.8 - 1.5$ GeV with the UV laser  
and $\sim 0.65 - 1.1$ GeV with the green one. The linear beam polarization
varied from $\sim 40\%$ at the lower energy limits up to $\sim 98\%$ at the upper ones.
The results obtained with two different types of runs were then used for cross-checks.   

Scattered photons were detected in a cylindrically symmetrical BGO
ball~\cite{bgo}. The ball
provided the detection of photons emitted at $\theta_{lab}=25 - 155^{\circ}$ with respect
to a beam axis. Recoil protons emitted at $\theta_{lab}\leq 25^{\circ}$ were detected 
in an assembly of forward detectors.
It consisted of two planar multi wire chambers,
a double hodoscope scintillator wall, and a lead-scintillator time-of-flight (TOF) wall~\cite{rw}.

The data analysis was similar to that used in the previous 
measurement on the neutron~\cite{comp}.
At first, the $\gamma p$ final states were
identified using the criterion of coplanarity, cuts on the
proton and photon missing masses, and comparing
the measured TOF and the polar angle of the recoil proton with the same quantities
calculated assuming  the $\gamma p \to \gamma p$ reaction.

The sample of the selected events was still populated by
events from the $\pi^0$ photoproduction.
Two types of the $\pi^0$ background
were taken into consideration:\\
i) Symmetric $\pi^0\to 2\gamma$ decays. The pion decays in two
photons of nearly equal energies. Being emitted in a narrow cone
along the pion trajectory,
such photons imitate a single-photon hit in the BGO ball;\\
ii) Asymmetric $\pi^0\to 2\gamma$ decays. One of the photons takes
the main part of the pion energy. It is emitted nearly along the
pion trajectory. Such photon and the recoil proton mimic Compton scattering.
The second photon is soft and is emitted into a
backward hemisphere relative to the pion track. Its energy depends
on the pion energy and may be as low as $6 - 10 $~MeV.

The symmetric events were efficiently rejected by analyzing the
distribution of energies deposited in crystals attributed to the
corresponding cluster in the BGO ball. The efficiency of this
rejection was verified in simulations and found to be $99\%$.

The asymmetric $\pi^0\to 2 \gamma$ decays were the major
problem. The GRAAL detector provides the low-threshold ($5$~MeV)
detection of photons in the nearly $4\pi$ solid angle. If one
(high-energy) photon would be emitted at backward angles, an the second (low-energy) photon 
could then be detected in the BGO ball or in the forward lead-scintillator wall.
This feature made it possible to suppress
the $\pi^0$ photoproduction.

For the further selection of events the
missing energy $E_{mis}$ was employed
\begin{equation}
E_{mis}=E_{\gamma}-E_{{\gamma^\prime}}-T_{p}(\theta_{p}),
\end{equation}
\noindent where $E_{\gamma}$ denotes the energy of the incoming
photon, $E_{\gamma^\prime}$ is the energy of the scattered photon,
and $T_{p}(\theta_{p})$ is the kinetic energy of the recoil
proton.

\begin{figure}
\vspace*{0.8cm}
\epsfverbosetrue\epsfxsize=8.2cm\epsfysize=6.7cm\epsfbox{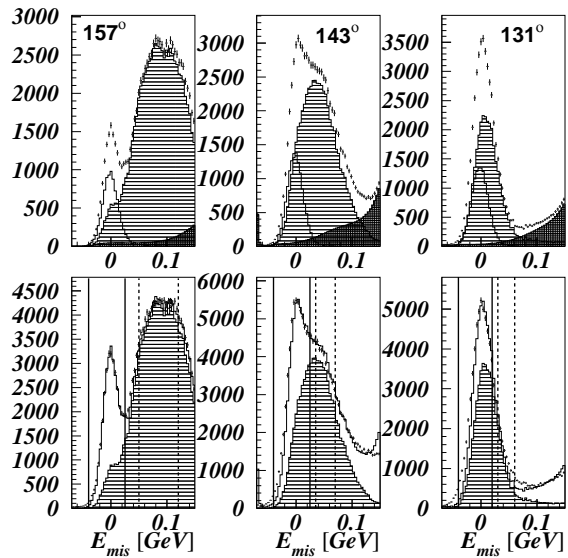}
\vspace*{-0.4cm}
\caption{Spectra of missing energy. Upper panels show the results of simulations. Solid
lines correspond to Compton events. Dashed areas are the events from $\gamma p \to \pi^0 p$.
Dark areas are the yields of other reactions.
Lower panels show the spectra obtained in experiment. Dashed areas are the estimated contamination 
of $\pi^0$ events. Solid lines indicate the cut used to select the mixture of Compton and $\pi^0$ events.
Dashed lines are the side-band cuts used to select $\pi^0$ events.}
\label{fig:me}
\vspace*{-0.4cm}
\end{figure}

The simulated spectra of the missing energy
are shown in the upper panels of Fig.~\ref{fig:me}. 
$\pi^0$ events form wide distributions. Compton events generate narrow peaks
centered around $E_{mis}=0$. The  events in this region 
belong to both Compton scattering and $\pi^0$ photoproduction. The contamination of events
from other reactions (mostly double neutral pion photoproduction) does not exceed $2\%$. At
larger $E_{mis}$ the spectra are dominated by $\pi^0$ events.

The Compton peak is clearly seen at $157^{\circ}$ (the angular bin $151 - 165^{\circ}$).
At these angles soft photons
from asymmetric $\pi^0$ decays are efficiently detected in either the BGO Ball
or in the forward shower wall. At more forward angles part of such photons
escapes out through the backward gap in the GRAAL detector.
The distributions  of Compton and $\pi^0$ events get closer being almost
unresolved at $131^{\circ}$ (the angular bin $122 - 137^{\circ}$). 

The experimental spectra (lower panels of Fig.~\ref{fig:me}) are quite similar
to the simulated ones. Solid lines show the cut $-0.04\leq E_{mis} \leq 0.025$.
This cut was used to select the mixture of Compton and $\pi^0$ events.
The events in the region above $E_{mis}=0.035$ GeV are mostly 
from $\pi^0$ photoproduction. Dashed lines in Fig.~\ref{fig:me} indicate side-band cuts.
These cuts select mostly $\pi^0$ events.

\begin{figure}
\vspace*{0.9cm}
\epsfverbosetrue\epsfxsize=9.2cm\epsfysize=7.2cm\epsfbox{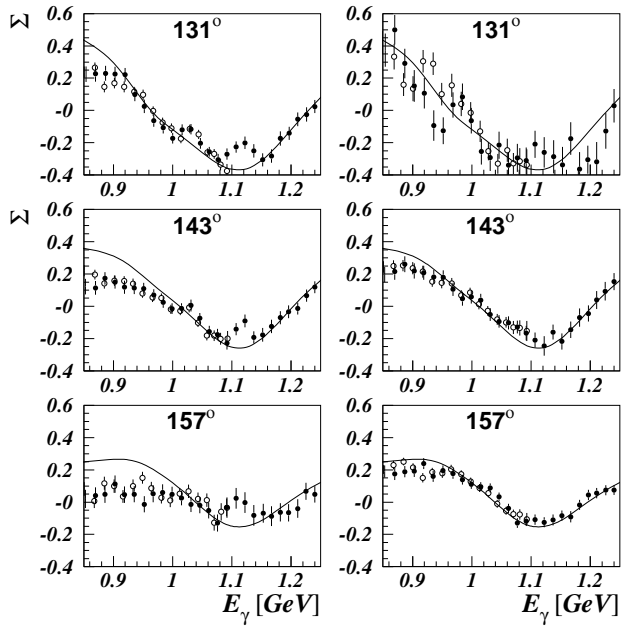}
\vspace*{-0.7cm} 
\caption{On the left: Beam asymmetry $\Sigma$ for the mixture of Compton and $\pi^0$
events. On the right: Beam asymmetry $\Sigma$   
obtained using side-band cuts (mostly $\pi^0$ events). Dark(open) circles are the results 
obtained with UV(green) laser.
Solid lines are the SAID SM11 solution for the $\gamma p \to \pi^0 p$ beam asymmetry.}
\vspace*{-0.6cm} 
\label{fig:ass1}
\end{figure}

Fig.~\ref{fig:ass1} shows the beam asymmetry $\Sigma$ of events selected using the main and side-band cuts.
The results obtained with the UV and green lasers are statistically independent. They are in good agreement.
The data points obtained with the side-band cuts (right panels of Fig.~\ref{fig:ass1})
are close to the SM11 solution of the SAID partial-wave analysis (PWA) for $\pi^0$ photoproduction.
The minor discrepancy is due to the contamination of Compton and other events.

The beam asymmetries of the mixture of Compton and 
$\pi^0$ events  (the main cut, left panels of Fig.~\ref{fig:ass1}) deviate
from the SM11 solution. There are two narrow structures which are not seen
with the side-band cuts.

%\begin{figure}
%\vspace*{1.0cm}
%\epsfverbosetrue\epsfxsize=7.2cm\epsfysize=5.8cm\epsfbox{ass1b.eps}
%\vspace*{-3.7cm} 
%\caption{Left: Beam asymmetry $\Sigma$ at $122 - 165^{\circ}$ obtained with the cut $-0.04\leq E_{mis}<0$ GeV.
%Middle: the same with the cut $0.\leq E_{mis}<0.025$ GeV. Right: The same with the cut
%$0.025\leq E_{mis}<0.05$ GeV. The legend for dark(open) circles is as in Fig.~\ref{fig:ass1}.}
%\vspace*{-0.6cm} 
%\label{fig:ass1b}
%\end{figure}

The validity of this observation was verified by means of different cuts of the missing energy 
%Fig.~\ref{fig:ass1b} shows the beam asymmetries 
in the overall angular range $122 - 165^{\circ}$, 
namely $-0.04\leq E_{mis}<0$ GeV, $0.\leq E_{mis}<0.025$ GeV, and $0.025\leq E_{mis}<0.05$ GeV.
The first two cuts selected the mixture of Compton and $\pi^0$ events. Both structures were seen with these cuts
and  dissapeared with the third one which selected mostly $\pi^0$ events.

The beam asymmetry shown in the left panels of Fig.~\ref{fig:ass1} is the combination  
of both Compton and $\pi^0$ beam asymmetries (the minor contribution of events from other reactions can be neglected)
\begin{equation}
\Sigma_{tot}=\alpha \Sigma_{comp}+(1-\alpha) \Sigma_{\pi^0}
\label{eq2}
\end{equation}
where $\alpha=\frac{N_{comp}}{N_{comp}+N_{\pi^0}}$ denotes the fraction of Compton events.

The contamination of $\pi^0$ events was determined by normalizing the simulated $\pi^0$
spectrum in the angular bin of $157^{\circ}$ to the experimental one 
in the region of the side-band cut (Fig.~\ref{fig:me}). Then the same normalization was used
to determine the $\pi^0$ contamination in two other angular bins.

The fraction of Compton events $\alpha$ varied from $\sim 90\%$ to $\sim 40\%$
at 0.85 to 1.25 GeV in the angular bin $157^{\circ}$, from $\sim 75\%$ to $\sim 35\%$ in the
angular bin $143^{\circ}$, and from from $\sim 60\%$ to $\sim 30\%$ in the
angular bin $131^{\circ}$. The $\pi^0$ beam asymmetry $\Sigma_{\pi^0}$ was taken from the SAID SM11 solution.
Then Compton beam asymmetry $\Sigma_{comp}$ was derived using Eq.~\ref{eq2}
%\begin{equation}
%\Sigma_{comp}=\frac{\Sigma_{exp}-(1-\alpha)\Sigma_{\pi^0}}{\alpha}
%\label{eq3}
%\end{equation}
in which the $\pi^0$ beam asymmetry $\Sigma_{\pi^0}$ was set equal to 
the SAID SM11 solution.

\begin{figure}
\vspace*{-0.1cm}
\epsfverbosetrue\epsfxsize=9.2cm\epsfysize=8.2cm\epsfbox{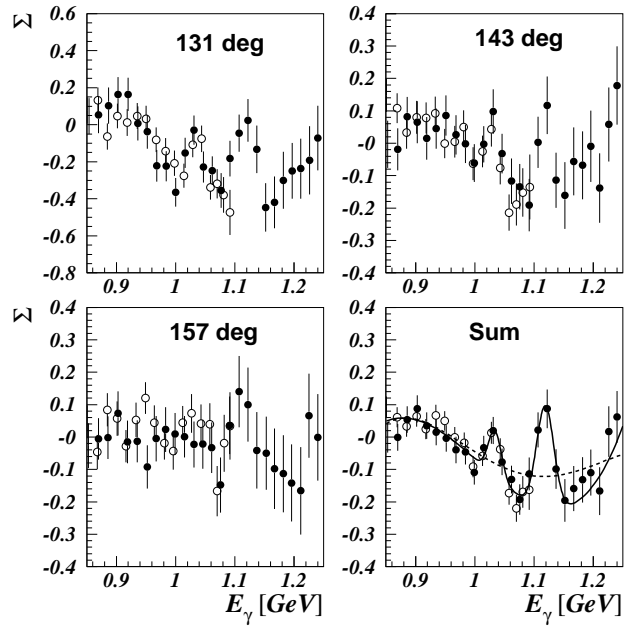}
\vspace*{-0.7cm} 
\caption{Beam asymmetry $\Sigma$ for Compton scattering on the proton.
Dark(open) circles are the results obtained with UV(green) laser.}
\vspace*{-0.5cm} 
\label{fig:ass3}
\end{figure}

The results are shown in Fig.~\ref{fig:ass3}. At the energies below $1$ GeV
the Compton beam asymmetry is close to 0. 
Above $1$ GeV there are two narrow structures. They are better pronounced at $131^{\circ}$
and almost degenerate at $157^{\circ}$. This is a typical trend for the beam asymmetry
$\Sigma$ which {\it a priori} approaches 0 at $180^{\circ}$.

Compton scattering was calculated 
by A.~L'vov {\it et al.}~\cite{lvov} on the base of dispersion relations. The range of model validity
is below $1$ GeV. No calculations of Compton scattering
at higher energies is available.
Because of lack of theoretical predictions the data was fit in a simple way:
the results from three angular bins were summed with weights proportional to inverse squares of their errors 
(lower right panel of Fig.~\ref{fig:ass3})
and fit either by the 4-order polynomial (the background hypothesis)
or by the 4-order polynomial-plus-two modified Breit-Wigner distributions (the background-plus-signal hypothesis).
The formula for the Breit-Wigner distributions
\begin{equation}
A_i\frac{(E_{\gamma}-E_{Ri})cos(\phi_i)+\Gamma_i sin(\phi_i)}{(E_{\gamma}-E_{Ri})^2+\frac{\Gamma_i ^2}{4}} , i=1,2
\label{eq4}
\end{equation}
was suggested in Ref.~\cite{am} to describe the interference between a narrow resonance and background. 
The mass centers of the distributions were extracted as $E_{R_1} = 1.036\pm 0.002$ GeV ($W_1 = 1.681$ GeV)
and $E_{R_2} = 1.119\pm 0.002$ GeV ($W_2 = 1.726$ GeV).  
The widths were $\Gamma_1 = 25\pm 10$ and $\Gamma_2 = 35\pm 12$ MeV (
$\Gamma_1 = 18\pm 6$ and $\Gamma_2 = 21\pm 7$ MeV in the units of the center-of-mass energy $W$). 
The $\chi$-squares of the fits were 75.7/39 (background hypothesis) and 29.7/31 (signal-plus background
hypothesis). The log likelihood ratio
of these two hypotheses ($\sqrt{2ln(L_{B+S}/L_{B}}$) corresponded
to the confidence level of $\approx 4.8 \sigma$.

The errors shown in Fig.~\ref{fig:ass3} are only statistical. The systematic uncertainty 
mainly originates from the determination of $\alpha$.  
One may see from the Eq.~\ref{eq2} that it less affects $\Sigma_{comp}$ 
if (i) $\alpha$ is large and (ii) $\Sigma_{comp}\approx\Sigma{\pi^0}$. 
This uncertainty mostly affects $\Sigma_{comp}$
in the regions of the observed structures. It results in the additional $\approx 20\%$ errors in 
the extraction of the amplitudes $A_i$ in Eq.~\ref{eq4}.

The observation of the narrow structure at $W \approx 1.68$ GeV correlates with the
previous results on $\eta$ photoproduction~\cite{gra,kru,kru2,kas,mainz1},
Compton scattering off the neutron~\cite{comp}, and $\eta$ photoproduction
on the proton~\cite{acta,jetp}. 
The second structure at $W \approx 1.73$ GeV was not seen in the mentioned experiments. 
However the modified SAID partial-wave analysis~\cite{arndt} hinted two narrow
$P_{11}$ resonances at $W=1.68$ GeV and  $W=1.73$ GeV. Both structures were also seen in the
preliminary data on $\pi N$ scattering by the EPECUR Collaboration~\cite{epe}. 
The preliminary evidence for the peak at $W=1.72$ GeV in $K\Lambda$ invariant mass
was reported by the STAR Collaboration~\cite{star} but remained unpublished.
The structure at $W\approx1.68$ GeV is one more challenge for the
explanation of the neutron anomaly in terms of the interference of well-known
resonances~\cite{int}. This hypothesis cannot explain all
experimental findings. 
%Moreover the recent revision of this idea through
%the flavor SU(3) symmetry~\cite{max1} raises more questions.

The energies $W\approx 1.68$ and $W\approx 1.73$ GeV correspond to the $K\Lambda$ and $\omega p$
photoproduction thresholds. This favors the cusp effect as an explanation of the 
neutron anomaly. Furthermore, a narrow step-like structure  was also observed at the
$K^*\Lambda$ threshold~\cite{schmid}.
On the other hand it still remains unclear as to (i) why this effect is not seen in $\pi N$
photoproduction, (ii) whether it could occur in Compton scattering, and (iii)
why the structure at $W\approx 1.72$ GeV is seen in Compton scattering
 and is not seen  in $\eta$ photoproduction on the neutron.

The observation of these structures may signal one or two narrow resonances. 
Their masses and width which stem
from our simple fit are $M_1=1.681\pm 0.002_{stat} \pm 0.005_{syst}$,  
$M_2=1.726\pm 0.002_{stat}\pm 0.005_{syst}$ GeV, $\Gamma_1=18\pm 6$ and $\Gamma_2=21\pm7 $ MeV.
The systematic errors $\Delta M$ are due to the accuracy of the calibration of the GRAAL
tagging system.

The decisive identification of both structures requires a common fit of Compton and
$\eta$ photoproduction data. Accurate calculations of Compton scattering are needed for that.
One particular task is to determine the waves and quantum numbers. 
Cusp is {\it a priori} an $S$-wave phenomenon. The Chiral Soliton Model predicts one exotic
$P_{11}$ state with the mass near $1.7$ GeV~\cite{dia}.

It is our pleasure to thank the staff of the European Synchrotron
Radiation Facility (Grenoble, France) for the stable beam
operation. This work was
supported by INFN Sezione di Catania and by High Energy Physics Department of Petersburg Nuclear Physics Institute.
Discussions with Profs. B.~Krusche, A.~L'vov, M.~Polyakov, and
H.~Schmieden were quite stimulating.

%%%%%%%%%%%%%%%%%%%%%%%%%%%%%%%%%%%%%%%%%%%%%%%%%%%%%%


\begin{thebibliography}{10}

\bibitem{gra}  V.~Kuznetsov {\it et al.},
  Phys.\ Lett.\  B {\bf 647}, 23 (2007).
\bibitem{kru}  I.~Jaegle {\it et al.}, Phys.\ Rev.\ Lett.\  {\bf 100}, 252002 (2008).
\bibitem{kru2}  I.~Jaegle {\it et al.}, Eur.Phys.J. A{\bf 47}, 89 (2011).
\bibitem{kas}  F.~Miyahara {\it et al.},
  Prog.\ Theor.\ Phys.\ Suppl.\  {\bf 168}, 90 (2007).
\bibitem{mainz1} D. Werthmuller {\it et al.}, Phys.Rev.Lett. {\bf }111 (2013) 23, 232001;
Phys.Rev. C{\bf 90}, 015205 (2014).
\bibitem{comp} V.Kuznetsov et al., Phys. Rev. C{\bf 83}, 022201 (2011).
\bibitem{az}    Y.~I.~Azimov, V.~Kuznetsov, M.~V.~Polyakov, and I.~Strakovsky,
  Eur.\ Phys.\ J.\  A {\bf 25}, 325 (2005).
\bibitem{tia}   A.~Fix, L.~Tiator and M.~V.~Polyakov,
  Eur.\ Phys.\ J.\  A {\bf 32}, 311 (2007).
\bibitem{kim}  K.~S.~Choi, S.~I.~Nam, A.~Hosaka and H.~C.~Kim,
  Phys.\ Lett.\  B {\bf 636}, 253 (2006).
\bibitem{arndt} R.~A.~Arndt, Ya.~I.~Azimov, M.~V.~Polyakov, I.~I.~Strakovsky, and R.~Workman,
  Phys.\ Rev.\  C {\bf 69}, 035208 (2004).
\bibitem{mart} T.Mart, Phys.Rev. D{\bf 83}, 094015 (2011).
\bibitem{cqm} N.~Isgur and G.~Karl, Phys.Rev. D~\textbf{18}, 4187, (1978);
N.~Isgur and G.~Karl, Phys. Lett. B\textbf{74}, 353, (1978); 
S.Capstick and W.Roberts, Prog. Part. Nucl. Phys. \textbf{45}, S241 (2000) .
\bibitem{dia}
 D.~Diakonov, V.~Petrov and M.~V.~Polyakov,
  Z.\ Phys.\  A {\bf 359}, 305 (1997).
\bibitem{max}M.~V.~Polyakov and A.~Rathke,
  Eur.\ Phys.\ J.\  A {\bf 18}, 691 (2003).
\bibitem{dia1}D.~Diakonov and V.~Petrov,
  Phys.\ Rev.\  D {\bf 69} 094011 (2004).
  \bibitem{michal}
  J.~R.~Ellis, M.~Karliner and M.~Praszalowicz,
  J. High. Energy Phys. {\bf 04} (2004) 002;
  M.~Praszalowicz,
  Acta Phys.\ Polon.\  B {\bf 35}, 1625 (2004);
  Ann. Phys.\  {\bf 13}, 709 (2004). 
\bibitem{mfa} D. Diakonov, V.Petrov, and A.Vladimirov, Phys.Rev. D{\bf 88}, 074030 (2013),
and references therein.
\bibitem{int} A.~V.~Anisovich {\it et al.}, Eur.\ Phys.\ J.\  A {\bf 41}, 13 (2009); 
V.~Shklyar, H.~Lenske and U.~Mosel, Phys.\ Lett.\  B {\bf 650}, 172 (2007); \
R. Shyam and O. Scholten, Phys. Rev. C {\bf 78} 065201 (2008); 
V.~Tryasuchev, Eur. Phys. J. A {\bf 50}, 120 (2014).
%\bibitem{kru_rev} B. Krusche, C. Wilkin, accepted for publication in Progress in Particle and Nuclear Physics,
%arXiv:1410.7680 [nucl-ex]. 
\bibitem{dor} M.~Doring and K.~Nakayama,
  Phys.\ Lett.\  B {\bf 683}, 145 (2010).
\bibitem{mainz2} E.F. McNicoll et al., Phys. Rev. C {\bf 82}, 035208 (2010).
\bibitem{acta}   V.~Kuznetsov {\it et al.},
  Acta Phys.\ Polon.\  B {\bf 39}, 1949 (2008).
\bibitem{jetp}  V.~Kuznetsov and M.~V.~Polyakov,
  JETP Lett.\  {\bf 88}, 347 (2008).
\bibitem{an1} It is worth to noting that the authors of Ref.~\protect\cite{ann} arrived at a different conclusion. Problems in their
in their analysis are discussed in detail in Ref.~\protect\cite{acta}.
\bibitem{ann} O.~Bartalini \textit{et al.}, Eur.\ Phys. \ J. A \textbf{33}, 169 (2007).
\bibitem{pi0} Detailed description of the GRAAL facility is available in
              O.~Bartalini \textit{et al.}, Eur. Phys. J. A\textbf{26}, 399 (2005).
\bibitem{bgo} F.~Ghio \textit{et al.}, Nucl. Inst. a. Meth. A\textbf{404}, 71 (1998).
\bibitem{rw}  V.~Kouznetsov {\it et al.}, Nucl. Inst. a. Meth. A\textbf{487}, 396 (2002).
\bibitem{lvov} A.~Lvov, V.~Petrun'kin, and M.~Shumacher, Phys. Rev. C{\bf 55}, 355, (1997), and A.~L'vov,
Private communication.
\bibitem{am} M.~Amarian, D.~Diakonov, M.~Polyakov, Phys. Rev. D{\bf 78}, 074003 (2008). 
\bibitem{epe} A.~Gridnev for the EPECUR Collaboration, PoS Hadron2013, 099 (2013).
\bibitem{star} S. Kabana for the STAR Collaboration,
    PoS of 20th Winter Workshop on Nuclear Dynamics, Trelawny
    Beach, Jamaica March 2003.
%\bibitem{max1} T.~Boika, V.~Kuznetsov, and M.V.~Polyakov, Submitted to Phys. Lett. B; arXiv:1411.4375 [nucl-th]. 
\bibitem{schmid} R.~Ewald {\it et al.}, Phys. Lett. B {\bf 713}, 180 (2012).

 
\end{thebibliography}
\end{document}